\documentclass[a4paper,11pt]{article}
\usepackage[english]{babel}
\usepackage{graphicx,color}
\usepackage{ifpdf}
\ifpdf
  \usepackage[pdftex]{hyperref}
\else
  \usepackage[hypertex]{hyperref}
\fi
\hypersetup{bookmarks=true, unicode=true, colorlinks=true, linkcolor=black, citecolor=black, filecolor=black, urlcolor=black}

\usepackage{amsmath,amssymb,amsfonts,bm,euscript,amscd,ae,amsthm,exscale,amsopn,amstext}
\usepackage{ifthen}
\usepackage[T1]{fontenc}
\usepackage[cp1251]{inputenc}
\usepackage[active]{srcltx}
\usepackage{setspace}

\newtheorem*{conclusion}{Conclusion}
\newtheorem*{proposition}{Proposition}
\newcommand{\tr}{\mathrm{Tr}}

\title{On the Residual Effective Potential within Global One-Dimensional Quantum Gravity}

\author{
\textbf{Lukasz Andrzej Glinka}\footnote{E-mail to: \href{mailto:laglinka@gmail.com}{\bf{\tt{laglinka@gmail.com}}}}}

\date{\today}

\begin{document}

\maketitle
\begin{abstract}
The global one-dimensional quantum gravity is the model of quantum gravity which arises from the global one-dimensionality conjecture within quantum general relativity, first considered by the author in 2010 and then in 2012. In this model the global dimension is a determinant of a metric of three-dimensional space embedded into an enveloping Lorentizan four-dimensional spacetime. In 2012, it has already been presented by the author that this model can be extended to any Lorentzian $D+1$-dimensional spacetime, where $D$ is a dimension of space, and resulting in the global one-dimensional model of a higher dimensional quantum gravity.

The purely quantum--mechanical part of this model is a minimal ef\/fective model within the quantum geometrodynamics, introduced by J.A. Wheeler and B.S. DeWitt in the 1960s, but the ef\/fective potential is manifestly different from the one considered by Wheeler \& DeWitt. Moreover, in our model the wave functionals solving the quantum gravity are one-variable smooth functions and, therefore, the troublesome mathematical technique of the Feynman functional integration present in the Hawking formulation of quantum gravity, is absent is this model, what makes it a mathematically consistent theory of quantum gravitation.

In this paper, we discuss in some detail a certain part of the global one-dimensional model already proposed in 2010, and then developed in 2012. The generalized functional expansion of the ef\/fective potential and the residual approximation, which describe the embedded spaces which are maximally symmetric three-dimensional Einstein's manifolds, whose lead to the Newton--Coulomb type potential in the quantum gravity model, are considered. Furthermore, scenarios related to few selected specific forms of the ef\/fective potential are suggested as physically interesting and discussed.\\

\end{abstract}

\newpage
\section{Introduction}
It was recently proposed by the author, Cf. the Refs. \cite{glinka1, glinka2}, to take into account the global one--dimensionality conjecture within quantum general relativity, where the global dimension is determinant of metric of a three-dimensional space embedded into four-dimensional spacetime. Actually, this conjecture leads to the non-trivial model of quantum gravity which dif\/fers from the standardly considered approaches \cite{qg}.

The quantum-mechanical part of the global one-dimensional quantum gravity can be considered separately as a nice mathematical theory having possibly interesting physical ramifications. In and of itself this fragment of the model is globally one-dimensional quantum mechanics describing $3+1$-decomposed solutions of the Einstein f\/ield equations of general relativity. In this theory quantum gravity is given by the one-dimensional Schr\"odinger equation, where the single dimension is the global dimension. The global quantum mechanics can be interpreted in terms of radial-type Schr\"odinger wave equation and, for this reason, it straightforwardly leads to the strict physical relation with atomic and nuclear physics.

In this paper few selected elements of the globally one-dimensional quantum mechanics are discussed in some detail. The generalized functional expansion of the ef\/fective potential and the residual approximation of the expansion, which corresponds to the embeddings being the maximally symmetric three-dimensional Einstein manifolds, whose the physical meaning is reconstruction the Newton--Coulomb type potential within the model of quantum gravity, are considered. Few possible mathematical scenarios with respect to the form of the ef\/fective potential are suggested as possibly interesting from the theoretical physics point of view.

The content of this paper is as follows. In the Section \ref{sec:1} the global one-dimensional model of quantum gravity is briefly discussed, and its quantum mechanical part is presented in some detail. We start, in the Subsection \ref{sub:1}, from the condensed presentation of the standard way resulting in the Wheeler--DeWitt equation of quantum geometrodynamics which is the standardly considered model of quantum gravity and is the core fundament of all presently considered approaches. Then, in the Subsection \ref{sub:2}, the idea of the global dimension, the resulting quantum mechanics, and the idea of the generalized functional expansion of the ef\/fective potential are presented. Next, in the Subsection \ref{sub:3}, the idea of the invariant global dimension is digressed only, with no continuation in the further part of the paper. In the Section \ref{sec:2}, the residual approximation of the ef\/fective potential, the role of maximally symmetric three-dimensional Einstein manifolds within the model of quantum gravity, and few conclusions having possibly interesting physical significance are discussed. The Section \ref{sec:3} is devoted to derivation of the ''geometric'' wave functionals of the global quantum mechanics in terms of the special functions, with respect to the selected situations given by the three types of boundary conditions discussed one by one in the Subsections \ref{sub:4}, \ref{sub:5}, and \ref{sub:6}. F\/inally, in the Section \ref{sec:5} we summarize briefly the paper.

\section{\mbox{Global One-Dimensional Quantum Gravity}}\label{sec:1}

\subsection{Standard Quantum Geometrodynamics}\label{sub:1}
In general relativity \cite{ein,hil} a Lorentzian (pseudo--Riemannian) \cite{rie} manifold $(M,g)$ with a metric $g_{\mu\nu}$ and a distance $ds^2=g_{\mu\nu}dx^\mu dx^\nu$, where $x^\mu$, $\mu=0,1,2,3$ is a coordinate system, characterized by the Christof\/fel symbols $\Gamma^\rho_{\mu\nu}$ and curvatures: Riemann--Christof\/fel  $R^\lambda_{\mu\alpha\nu}$, Ricci $R_{\mu\nu}$, Ricci scalar $R$
\begin{eqnarray}
\Gamma^\rho_{\mu\nu}&=&\dfrac{1}{2}g^{\rho\sigma}\left(g_{\mu\sigma,\nu}+g_{\sigma\nu,\mu}-g_{\mu\nu,\sigma}\right),\nonumber\\
R^\lambda_{\mu\alpha\nu}&=&\Gamma^\lambda_{\mu\nu,\alpha}-\Gamma^\lambda_{\mu\alpha,\nu}+\Gamma^\lambda_{\sigma\alpha}\Gamma^\sigma_{\mu\nu}-\Gamma^\lambda_{\sigma\nu}\Gamma^\sigma_{\mu\alpha},\\
R_{\mu\nu}&=&R^\lambda_{\mu\lambda\nu},\nonumber\\
R&=&g^{\kappa\lambda}R_{\kappa\lambda},\nonumber
\end{eqnarray}
is a spacetime being a solution of the Einstein f\/ield equations
\begin{equation}\label{feq}
G_{\mu\nu}+\Lambda g_{\mu\nu}=\kappa T_{\mu\nu},
\end{equation}
where $\kappa$ is the Einstein constant, $G_{\mu\nu}\equiv R_{\mu\nu}-\dfrac{1}{2}Rg_{\mu\nu}$ is the Einstein tensor, $\Lambda$ is cosmological constant, and $T_{\mu\nu}$ is stress-energy tensor of Matter f\/ields which arises through the variational principle \cite{pal} applied to the Einstein--Hilbert action modif\/ied through the York--Gibbons--Hawking boundary term \cite{ygh}
\begin{equation}\label{eh0}
S[g]\!=\!\int_{M}d\mu_g\left\{-\dfrac{R}{2\kappa}+\dfrac{\Lambda}{\kappa}+\mathcal{L}\right\}-\dfrac{1}{\kappa}\int_{\partial M}d\mu_hK\quad,
\end{equation}
where $K$ is the Gauss scalar curvature of a spacelike boundary $(\partial M,h)$, $\mathcal{L}$ is the Lagrangian of Matter fields, and $d\mu_g=d^4x\sqrt{-g}$, $d\mu_h=d^3x\sqrt{h}$ are the invariant measures of four-dimensional spacetime and embedded three-dimensional space, respectively.

Application of the Nash embedding theorem \cite{nash} allows to justify the $3+1$ decomposition of spacetime metric \cite{adm}
\begin{eqnarray}\label{dec}
g_{\mu\nu}=\left[\begin{array}{cc}-N^2+N^iN_i&N_j\\N_i&h_{ij}\end{array}\right]\quad,\quad h_{ik}h^{kj}=\delta_i^j\quad,
\end{eqnarray}
where $N$, $N_i$, $h_{ij}$ are respectively called the lapse function, the shift vector, and embedded space metric, $N^i=h^{ij}N_j$ is the intrinsic covariant shift vector. The $3+1$ decomposition transforms the action functional (\ref{eh0}) into the Hamiltonian form
\begin{eqnarray}\label{gd}
S[g]=\int dt\int_{\partial M} d^3x\left\{\pi_\phi \dot{\phi}+\pi\dot{N}+\pi^i\dot{N_i}+\pi^{ij}\dot{h}_{ij}-NH-N_iH^i\right\},
\end{eqnarray}
where $\phi$ symbolizes Matter f\/ields, and nontrivial $\pi$'s, $H$, $H^i$ are \cite{gau}
\begin{eqnarray}
\pi^{ij}&=&-\dfrac{\sqrt{h}}{2\kappa}\left(K^{ij}-Kh^{ij}\right)\quad,\nonumber\\
H&=&\dfrac{\sqrt{h}}{2\kappa}\left\{{^{(3)}\!R}+K^2-K_{ij}K^{ij}-2\Lambda-2\kappa\varrho\right\},\label{con}\\
H^i&=&2\pi^{ij}_{~;j},\nonumber
\end{eqnarray}
where ${^{(3)}\!R}$ is the Ricci scalar of an embedding, $\varrho=n^{\mu}n^{\nu}T_{\mu\nu}$ is the Matter f\/ields energy related to the normal vector f\/ield $n^\mu=[1/N,-N^i/N]$. Extrinsic curvature $K_{ij}$, where $\tr K_{ij}\equiv K$, is constrained by the equality
\begin{equation}
  \dot{h}_{ij}=2\left(NK_{ij}+N_{(i|j)}\right),\label{con0}
\end{equation}
where $N_{(i|j)}$ is symmetrized intrinsic covariant derivative of the shift. In accordance with DeWitt's foundational considerations \cite{dew}, $H^i$ are the generators of the spatial dif\/feomorphisms $\widetilde{x}^i=x^i+\delta x^i$
\begin{eqnarray}
i\left[h_{ij},\int_{\partial M}H_{a}\xi^a d^3x\right]&=&-h_{ij,k}\xi^k-h_{kj}\xi^{k}_{~,i}-h_{ik}\xi^{k}_{~,j},\\
i\left[\pi^{ij},\int_{\partial M}H_{a}\xi^a d^3x\right]&=&-\left(\pi^{ij}\xi^k\right)_{,k}+\pi^{kj}\xi^{i}_{~,k}+\pi^{ik}\xi^{j}_{~,k},\nonumber
\end{eqnarray}
where $H_i=h_{ij}H^j$. Time-preservation \cite{dir} of the primary constraints $\pi\approx0$ and $\pi^i\approx0$ leads to secondary ones - scalar and vector constraints
\begin{eqnarray}
H&\approx&0,\\
H^i&\approx&0,\nonumber
\end{eqnarray}
which create nontrivial f\/irst-class type constraints algebra \cite{dew}
\begin{eqnarray}
i\left[\int_{\partial M}H\delta x_1d^3x,\int_{\partial M}H\delta x_2d^3x\right]&=&\int_{\partial M}H^a\left(\delta x_{1,a}\delta x_2-\delta x_1\delta x_{2,a}\right)d^3x,\nonumber\\
i\left[H_i(x),H_j(y)\right]&=&\int_{\partial M}H_{a}c^a_{ij}d^3z,\label{com2}\\
i\left[H(x),H_i(y)\right]&=&H\delta^{(3)}_{,i}(x,y),\nonumber
\end{eqnarray}
where
\begin{equation}
c^a_{ij}=c^a_{ij}[x,y,z]=\delta^a_i\delta^b_j\delta^{(3)}_{,b}(x,z)\delta^{(3)}(y,z)-(i\leftrightarrow j,x\leftrightarrow y),
\end{equation}
are structure constants of the dif\/feomorphism group, and all Lie's brackets of $\pi$'s and $H$'s vanish. Scalar constraint determines dynamics, vector one merely reflects dif\/feoinvariance. Making use of the conjugate momenta, first formula in (\ref{con}), the scalar constraint transforms into the Hamilton--Jacobi type equation
\begin{equation}\label{con}
H=G_{ijkl}\pi^{ij}\pi^{kl}+\sqrt{h}\left(^{(3)}R-2\Lambda-2\kappa\varrho\right)\approx0\quad,
\end{equation}
where
\begin{equation}
G_{ijkl}\equiv\dfrac{1}{2\sqrt{h}}\left(h_{ik}h_{jl}+h_{il}h_{jk}-h_{ij}h_{kl}\right),
\end{equation}
is the DeWitt metric on the Wheeler superspace, a factor space of all $C^\infty$ Riemannian metrics on $\partial M$, and a group of all $C^\infty$ dif\/feomorphisms of $\partial M$ that preserve orientation \cite{sup}. The Dirac--Faddeev primary canonical quantization method \cite{dir,fad} in the present case has the form
\begin{eqnarray}\label{dpq}
i\left[\pi^{ij}(x),h_{kl}(y)\right]&=&\dfrac{1}{2}\left(\delta_{k}^{i}\delta_{l}^{j}+\delta_{l}^{i}\delta_{k}^{j}\right)\delta^{(3)}(x,y),\nonumber\\
i\left[\pi^i(x),N_j(y)\right]&=&\delta^i_j\delta^{(3)}(x,y),\\
 i\left[\pi(x),N(y)\right]&=&\delta^{(3)}(x,y).\nonumber
\end{eqnarray}
The solutions or rather representations of the momenta operators satisfying these canonical commutation relations is the question of choice. In quantum geometrodynamics, the Wheeler metric representation is usually taken into account. In such a representation the momenta operators are analogous to the momentum operator in quantum mechanics
\begin{eqnarray}
\pi&=&-i\dfrac{\delta}{\delta N},\nonumber\\
\pi^i&=&-i\dfrac{\delta}{\delta N_i},\\
\pi^{ij}&=&-i\dfrac{\delta}{\delta h_{ij}},\nonumber
\end{eqnarray}
and applied to the Hamiltonian constraint (\ref{con}) lead to the Wheeler--DeWitt equation \cite{whe, dew}
\begin{equation}\label{wdw}
\left\{2\kappa G_{ijkl}\dfrac{\delta^2}{\delta h_{ij}\delta h_{kl}}+\dfrac{\sqrt{h}}{2\kappa}\left(~{^{(3)}\!R}-2\Lambda-2\kappa\varrho\right)\right\}\Psi[h_{ij},\phi]=0\quad,
\end{equation}
and other f\/irst class constraints
\begin{eqnarray}
  \pi\Psi[h_{ij},\phi]&=&0,\nonumber\\
  \pi^i\Psi[h_{ij},\phi]&=&0,\\
  H^i\Psi[h_{ij},\phi]&=&0,\nonumber
\end{eqnarray}
merely reflect dif\/feoinvariance, and are not important for this model. Since about 60 years classical and quantum aspects of the Wheeler--DeWitt geometrodynamics have studied widely (Cf. \emph{e.g.} the books and collective volumes in Ref. \cite{qg}, and individual papers, Cf. \emph{e.g.} the Refs. \cite{can1,can2,can3,can4,can6,can7,can8,can9,can10,haw}).

\subsection{The Global Dimension}\label{sub:2}
The global one--dimensionality conjecture within quantum general relativity \cite{glinka1,glinka2} assumes the effective physical role of the determinant of a three-dimensional space embedded into a four-dimensional spacetime
\begin{equation}
  h=\det h_{ij}=\dfrac{1}{3}\varepsilon^{ijk}\varepsilon^{abc}h_{ia}h_{jb}h_{kc}\quad,
\end{equation}
where $\varepsilon^{ijk}$ is the Levi-Civita density, and takes into account the following situation as physically interersting
\begin{eqnarray}\label{GOD}
  \phi(x)&\rightarrow&\phi[h],\nonumber\\
  \varrho(\phi)&\rightarrow&\varrho[h],\\
  \Psi[h_{ij},\phi]&\rightarrow&\Psi[h].\nonumber
\end{eqnarray}
In other words, it replaces the functional dependence on the space metric elements and Matter fields through the function dependence on the determinant of a space metric. Applying the transformation of variables $h_{ij}\rightarrow h$ to the Wheeler--DeWitt equation (\ref{wdw}), one must change the functional dif\/ferentiation with respect to $h_{ij}$ by the classical differentiation with respect to $h$. It can be simply done through application of the Jacobi rule for dif\/ferentiation of the determinant of four-dimensional metric $\delta g = gg^{\mu\nu}\delta g_{\mu\nu}$ which leads to the result
\begin{equation}
  \delta h = hh^{ij}\delta h_{ij}\longrightarrow\dfrac{\delta}{\delta h_{ij}}=hh^{ij}\dfrac{\delta}{\delta h}.
\end{equation}
More detailed explanation can be found in the Ref. \cite{glinka1}.

Making few very elementary algebraic manipulations, one obtains from the equation (\ref{wdw}) the globally one-dimensional quantum mechanics
\begin{equation}\label{kgf}
\left(\dfrac{\delta^2}{\delta{h^2}}+V_{ef\/f}[h]\right)\Psi[h]=0.
\end{equation}
Here $V_{ef\/f}[h]$ is the ef\/fective potential
\begin{equation}\label{eff}
  V_{ef\/f}[h]\equiv V_G[h]+V_C[h]+V_M[h]\quad,
\end{equation}
which is a simple algebraic sum of the three fundamental energetic constituents
\begin{eqnarray}\label{pots}
V_G[h]&=&-\dfrac{1}{6\kappa^2}\dfrac{{^{(3)}\!R}}{h},\nonumber\\
V_C[h]&=&\dfrac{1}{3\kappa^2}\dfrac{\Lambda}{h},\\
V_M[h]&=&\dfrac{1}{3\kappa}\dfrac{\varrho[h]}{h},\nonumber
\end{eqnarray}
related to pure geometry of an embedded three-dimensional space ($G$), cosmological constant ($C$), and Matter f\/ields ($M$).

On the one hand, identif\/ication of the ef\/fective potential with the square of mass of the boson $V_{ef\/f}[h]\equiv m^2$ expresses the model of quantum gravity (\ref{kgf}) as the classical theory of massive bosonic f\/ield $\Psi[h]$. The construction of quantum f\/ield theory by the method of the static Fock reper of creators and annihilators, and related thermodynamics of quantum states can be also done elementary. This part of the model of quantum gravity was discussed in the 2012' monograph by the author \cite{glinka1}, and is not the main motive of this paper.

On the other hand, one can take into account the non-relativistic interpretation of the one--dimensional evolution (\ref{kgf}), and treat the received global quantum gravity model as the ef\/fective one-dimensional Schr\"odinger quantum mechanics with a certain selected effective potential being a functional of determinant of a three-dimensional embedding. In the spirit of this approach the potential $V_{ef\/f}[h]$ has intriguing meaning -- (\ref{eff}) is the equality between ''ef\/fective physics'', which can be constructed by other type considerations, and three basic constituents related to an embedding space -- ''geometric'', ''cosmological'', and ''material''.

Let us assume that the concrete form of $V_{ef\/f}$ is f\/ixed. Then one can express the Ricci scalar of a three-dimensional embedding as follows
\begin{equation}\label{kgf0}
{^{(3)}\!R}=2\kappa\varrho[h]+2\Lambda-6\kappa^2hV_{ef\/f}[h],
\end{equation}
and, therefore, the two last terms in the brackets can be treated as the dark energy density contribution
\begin{equation}
\varrho_{DM}[h]=\dfrac{\Lambda}{\kappa}-3\kappa hV_{ef\/f}[h].
\end{equation}
One can list several typical examples of the physical scenarios within the global one-dimensional model of quantum gravity, with respect to the concretely fixed form of the effective gravitational potential $V_{eff}$.
\begin{enumerate}
\item The case of the constant non-vanishing effective gravitational potential $V_{eff}=V_c\neq0$. In such a situation, the Ricci scalar curvature of the embedded space and the global one-dimensional quantum gravity are given by the equations
\begin{eqnarray}
&&{^{(3)}}R=2\Lambda+2\kappa \varrho-6\kappa^2hV_c,\label{kgf0a}\\
&&\left(\dfrac{\delta^2}{\delta{h^2}}+V_c\right)\Psi_c[h]=0,
\end{eqnarray}
where $\Psi_c[h]$ is a wave functional related to $V_{eff}=V_c$.
\item The case of the trivial effective gravitational potential $V_{eff}=0$. In such a situation, the three-dimensional Ricci scalar curvature and the global one-dimensional quantum gravity are
\begin{eqnarray}
&&{^{(3)}\!R}=2\Lambda+2\kappa \varrho,\label{kgf1}\\
&&\dfrac{\delta^2}{\delta{h^2}}\Psi_0[h]=0,
\end{eqnarray}
where $\Psi_0$ is a “free” wave functional related to $V_{eff}=0$.
\item The case when the sum of the geometric and cosmological contributions is identically vanishing $V_G+V_C=0$, but the material contribution does not vanish identically $V_{M}\neq0$. In such a situation, the three-dimensional Ricci scalar curvature and the global one-dimensional quantum gravity are
\begin{eqnarray}
&&{^{(3)}}R=2\Lambda,\label{kgf2}\\
&&\left(\dfrac{\delta^2}{\delta{h^2}}-\dfrac{1}{6\kappa^2}\dfrac{2\kappa }{h}\varrho[h]\right)\Psi_M[h]=0,
\end{eqnarray}
where $\Psi_M$ is a “material” wave functional related to the material contribution $V_M\neq0$.
\item The case when the sum of the geometric and material contributions is identically vanishing $V_G+V_M=0$, while in general the cosmological contribution is non-trivial $V_{C}\neq0$. In such a situation, the Ricci scalar curvature of the embedded space and the global one-dimensional quantum gravity are respectively
\begin{eqnarray}
&&{^{(3)}\!R}=2\kappa \varrho,\label{kgf4}\\ &&\left(\dfrac{\delta^2}{\delta{h^2}}+\dfrac{1}{6\kappa^2}\dfrac{2\Lambda}{h}\right)\Psi_C[h]=0.
\end{eqnarray}
Here $\Psi_C$ is the “cosmological” wave functional related to the cosmological contribution $V_C\neq0$.
\item The case when the sum of the cosmological and material contributions is identically vanishing $V_C+V_M=0$, whereas the geometric contribution is non-zero $V_{G}\neq0$. In such a situation, the energy density of Matter fields and the global one-dimensional quantum gravity are given by the equations
\begin{eqnarray}
&&\varrho=-\dfrac{\Lambda}{\kappa },\label{kgf3}\\ &&\left(\dfrac{\delta^2}{\delta{h^2}}-\dfrac{1}{6\kappa^2}\dfrac{{^{(3)}\!R}}{h}\right)\Psi_G[h]=0,
\end{eqnarray}
where $\Psi_G$ is the “geometric” wave functional related to the geometric contribution $V_G\neq0$.
\item A more general explicit form of the effective gravitational potential can be constructed in the spirit of complex analysis. Let us consider \emph{ad hoc} functional generalization of the Laurent series expansion in the global dimension $h$ of the effective gravitational potential $V_{eff}[h]$
    \begin{equation}\label{exp}
  V_{eff}[h]=\sum_{-\infty}^{\infty}a_{n}\left(h-h_0\right)^n,
   \end{equation}
    in an infinitesimal neighborhood, \emph{i.e.} in a one-sphere (circle) of radius $h_\epsilon$, of any \emph{ad hoc} fixed initial value $h_0$ of the global dimension
    \begin{equation}
    C(h_\epsilon)=\left\{h:|h-h_0|<h_{\epsilon}\right\}.
    \end{equation}
    The numbers $a_n$ are the series coefficients determined by the classical functional integral
\begin{equation}\label{calka}
  a_n=\dfrac{1}{2\pi i}\int_{C(h_\epsilon)}\dfrac{V_{eff}[h]}{\left(h-h_0\right)^{n+1}}\delta h,
\end{equation}
which is straightforward functional generalization of the Cauchy integral formula with the Lebesgue--Stieltjes measure -- the classical functional Radon measure $\delta h$.

 Let us take into considerations the most general choice of $h_0$ which similarly to $h$ is assumed to be a complex number. In such a situation, the Ricci scalar curvature of a three-dimensional embedded space takes the following form
\begin{equation}\label{eq2}
  {^{(3)}\!R}=2\Lambda+2\kappa \varrho-6\kappa^2\sum_{-\infty}^{\infty}b_{n}(h-h_0)^{n},
\end{equation}
where $b_n$ is the combined series coefficient
\begin{equation}\label{been}
  b_n=a_{n-1}+h_0a_n=\dfrac{1}{2\pi i}\int_{C(h_\epsilon)}\dfrac{h}{\left(h-h_0\right)^{n+1}}V_{eff}[h]\delta h,
\end{equation}
and the global one-dimensional model of quantum gravity is then defined by the equation
\begin{equation}\label{eq1}
  \left(\dfrac{\delta^2}{\delta{h^2}}+\sum_{-\infty}^{\infty}a_{n}(h-h_0)^n\right)\Psi[h]=0.
\end{equation}
Making use of the triangle inequality one can write
\begin{equation}
  |b_n|\leqslant|a_{n-1}|+|h_0||a_n|
\end{equation}
so that it can be deduced straightforwardly that
\begin{equation}\label{been}
  \dfrac{|b_n|}{|a_n|}\leqslant\dfrac{|a_{n-1}|}{|a_n|}+|h_0|.
\end{equation}
Applying the well-known inequality for any Riemann integral
\begin{equation}
  \left|\int{f}\right|\leqslant\int|f|,
\end{equation}
where $f$ is considered as the Riemann-integrable function and the integral is considered as defined, to the coefficients $a_n$ and $b_n$ one obtains the inequality
\begin{eqnarray}
|a_n|&\leqslant&\dfrac{1}{h_\epsilon^{n+1}}\dfrac{1}{2\pi}\int_{C(h_\epsilon)}\left|V_{eff}\right|\delta{h}\leqslant\dfrac{1}{h_\epsilon^{n+1}}|a_{-1}|,
\end{eqnarray}
where $a_{-1}$ is the residue of the effective gravitational potential at the fixed point $h=h_0$. This residue is determined by the straightforward functional generalization of the Cauchy integral formula
\begin{equation}
  a_{-1}=\mathrm{Res}(V_{eff},h_0)=\dfrac{1}{2\pi i}\int_{C(h_\epsilon)}{V}_{eff}\delta{h},
\end{equation}
where $C(h_\epsilon)$ traces out circle around the fixed point $h_0$ in counter-clockwise manner on the punctured disk $D=\left\{z:0<|h-h_0|<R\right\}$. If the center of the circle $h=h_0$ is a pole of order $n$, then the residue is defined by the simple limiting procedure
\begin{equation}
\mathrm{Res}(V_{eff},h_0)=\dfrac{1}{\Gamma(n)}\lim_{h\rightarrow h_0}\dfrac{\delta^{n-1}}{\delta{h}^{n-1}}\left((h-h_0)V_{eff}\right).
\end{equation}

Making use of the following reasoning
\begin{equation}
  |a_n|=\left|\dfrac{1}{2\pi i}\int_{C(h_\epsilon)}\dfrac{V_{eff}}{(h-h_0)^{n+1}}\right|
  \leqslant\dfrac{1}{h_\epsilon}\left|\dfrac{1}{2\pi i}\int_{C(h_\epsilon)}\dfrac{V_{eff}}{(h-h_0)^{n}}\right|=\dfrac{1}{h_\epsilon}|a_{n-1}|,
\end{equation}
one sees that for any $n$ the inequality holds
\begin{equation}
  \dfrac{|a_{n-1}|}{|a_n|}\geqslant{h}_\epsilon,
\end{equation}
and hence the inequality (\ref{been}) can be rewritten as
\begin{equation}
  \dfrac{|b_n|}{|a_n|}\geqslant{h}_\epsilon+|h_0|.\label{inequalit1}
\end{equation}
In the light of the triangle inequality one can write
\begin{equation}
  |b_{n+1}|=|a_n+h_0a_{n+1}|\leqslant|a_n|+|h_0||a_{n+1}|,
\end{equation}
and because of the relation
\begin{equation}
\dfrac{|a_{n+1}|}{|a_n|}\leqslant\dfrac{1}{h_\epsilon},
\end{equation}
one receives the following relation
\begin{equation}
\dfrac{|b_{n+1}|}{|a_n|}\leqslant1+\dfrac{|h_0|}{h_\epsilon}.
\end{equation}
Finally, application of the inequality (\ref{inequalit1}) in the equivalent form
\begin{equation}
  \dfrac{|a_n|}{|b_n|}\leqslant\dfrac{1}{{h}_\epsilon+|h_0|}.\label{inequalit2}
\end{equation}
lead us to the following upper bound
\begin{equation}\label{boundy1}
  \dfrac{|b_{n+1}|}{|a_n|}\dfrac{|a_n|}{|b_n|}=\dfrac{|b_{n+1}|}{|b_n|}\leqslant\dfrac{1}{{h}_\epsilon+|h_0|}\left(1+\dfrac{|h_0|}{h_\epsilon}\right).
\end{equation}

Another bound for $\dfrac{|b_{n+1}|}{|b_n|}$ can be obtained as follows. In the light of the definition (\ref{been}) one can write
\begin{equation}
  a_n=\dfrac{b_n-a_{n-1}}{h_0},
\end{equation}
and, consequently, one can deduce the recursive relation for the coefficients $b_n$
\begin{equation}
  b_{n+1}=\dfrac{b_n-a_{n-1}}{h_0}+h_0a_{n+1},
\end{equation}
which after simple algebraic manipulations leads to the relation
\begin{equation}
  h_0b_{n+1}+a_{n-1}=h_0a_{n+1}+b_n.
\end{equation}
This equation can be rewritten as
\begin{equation}
  1=\left|\dfrac{h_0b_{n+1}}{h_0a_{n+1}+b_n}+\dfrac{a_{n-1}}{h_0a_{n+1}+b_n}\right|,
\end{equation}
which after taking into account the triangle inequality gives the relation
\begin{equation}
\left|\dfrac{h_0b_{n+1}}{h_0a_{n+1}+b_n}+\dfrac{a_{n-1}}{h_0a_{n+1}+b_n}\right|\leqslant\left|\dfrac{h_0b_{n+1}}{h_0a_{n+1}+b_n}\right|+\left|\dfrac{a_{n-1}}{h_0a_{n+1}+b_n}\right|,
\end{equation}
which leads to the conclusion
\begin{equation}
  |h_0a_{n+1}+b_n|\leqslant|h_0||b_{n+1}|+|a_{n-1}|.
\end{equation}
Once again, making use of the triangle inequality one has
\begin{equation}
  |h_0a_{n+1}+b_n|\leqslant|h_0||a_{n+1}|+|b_n|,
\end{equation}
and, consequently,
\begin{equation}
|h_0||b_{n+1}|-|b_n|\leqslant|h_0||a_{n+1}|-|a_{n-1}|.
\end{equation}
This inequality can be rewritten in the following form
\begin{equation}
|h_0|\dfrac{|b_{n+1}|}{|a_n|}-\dfrac{|b_n|}{|a_n|}\leqslant|h_0|\dfrac{|a_{n+1}|}{|a_n|}-\dfrac{|a_{n-1}|}{|a_n|},
\end{equation}
or more conveniently
\begin{equation}
|h_0|\dfrac{|b_{n+1}|}{|b_n|}-1\leqslant\dfrac{|a_n|}{|b_n|}\left(|h_0|\dfrac{|a_{n+1}|}{|a_n|}-\dfrac{|a_{n-1}|}{|a_n|}\right).
\end{equation}
In the light of the inequality (\ref{inequalit2}) and the relation
\begin{equation}
  |h_0|\dfrac{|a_{n+1}|}{|a_n|}-\dfrac{|a_{n-1}|}{|a_n|}\leqslant\dfrac{|h_0|}{{h}_\epsilon}-{h}_\epsilon,
\end{equation}
one obtains finally the following bound
\begin{equation}
\dfrac{|b_{n+1}|}{|b_n|}\leqslant\dfrac{1}{|h_0|}\left(1+\dfrac{1}{h_\epsilon+|h_0|}\left(\dfrac{|h_0|}{h_\epsilon}-h_\epsilon\right)\right),
\end{equation}
which taken together with the previous result (\ref{boundy1}) allows to deduce the inequality for the fixed point
\begin{equation}
|h_0|(|h_0|-1)\geqslant0,
\end{equation}
which can be resolved immediately and gives the consistency condition for the fixed point $h_0$
\begin{equation}
  |h_0|\in\{0\}\cup[1,\infty),\label{h0cond}
\end{equation}
and in itself is a non-trivial solution of the initial data problem.
\end{enumerate}
Naturally, there is plenty of other possibilities for the choice of a concrete form of the effective gravitational potential $V_{eff}[h]$. However, in the next section we shall discuss only a particular situation.

\subsection{The invariant global dimension}\label{sub:3}
Let us note that in general the global one-dimensional quantum mechanics (\ref{kgf}) cane be transformed by the second change of variables
\begin{equation}\label{kgfrx}
h\rightarrow \xi[h],
\end{equation}
where $\xi[h]$ is any functional in the global dimension $h$. In this case one can rewrite the global one-dimensional wave equation (\ref{kgf}) in the form
\begin{equation}\label{kgfr}
\left\{\left(\dfrac{\delta \xi[h]}{\delta h}\right)^2\dfrac{\delta^2}{\delta{\xi[h]^2}}+V_{ef\/f}\left[\xi[h]\right]\right\}\Psi\left[\xi[h]\right]=0,
\end{equation}
and if the coef\/f\/icient $\left(\dfrac{\delta \xi[h]}{\delta h}\right)^2$ does not vanish identically (or the transformation (\ref{kgfrx}) is non singular) then the equation (\ref{kgfr}) can be rewritten as
\begin{equation}\label{kgfr}
\left\{\dfrac{\delta^2}{\delta{\xi^2}}+V[\xi]\right\}\Psi\left[\xi\right]=0,
\end{equation}
where the new potential $V[\xi]$ is scaled ef\/fective potential $V_{ef\/f}$ expressed by the new dimension $\xi$
\begin{equation}
  V[\xi]=\Omega^2[\xi[h]]V_{ef\/f}\left[\xi[h]\right]\quad,\quad\Omega[\xi[h]]=\left(\dfrac{\delta \xi[h]}{\delta h}\right)^{-1}.
\end{equation}
One sees easily that the following choice of the ''gauge'' $\xi[h]$
\begin{equation}\label{lg}
  \xi[h]\equiv h,
\end{equation}
transforms the quantum mechanics (\ref{kgf}) into itself. The choice of the transformation of variables in the form (\ref{lg}) is the simplest transformation of the kind $h_{ij}\rightarrow \xi[\det h_{ij}]$ within the Wheeler--DeWitt theory. Other, more advanced propositions, can be generated directly from this basic case, and should be justif\/ied by some rational arguments. Let us choose the transformation of variables in the form
\begin{equation}\label{invariant}
\xi[h]=\sqrt{h}.
\end{equation}
It is clear that this selection can be justif\/ied by the fact that $\sqrt{h}$ is the invariant volume element on an embedding with assumption that $h>0$\footnote{for $h<0$, $\sqrt{h}$ should be replaced by $\sqrt{|h|}$, and we do not lose generality.}. In this manner the dimension $\sqrt{h}$ has an invariant nature. The choice (\ref{invariant}) yields the equation (\ref{kgfr}) with the following modif\/ied ef\/fective potential
\begin{equation}
V[\xi]=4\xi^2V_{ef\/f}[\xi].
\end{equation}
Moreover, the singularity $\dfrac{1}{h}$ evidently vanishes, but actually causes that $V[\xi]$ must be studied with respect to the new ''invariant'' dimension $\xi$.

The very good point of reference in searching for the dimension $\xi$ is the normalization condition of the Schr\"odinger quantum mechanics, which for the considered situation takes the form of the Lebesgue--Stieltjes/Radon integral
\begin{equation}\label{norm}
  \int_{\Omega(h_I,h)}\left|\Psi\left[\xi[h]\right]\right|^2\delta \xi[h]=1,
\end{equation}
where $\Omega(h_I,h)$ is some region of integrability in a space of all three-dimensional embeddings with metric $h_{ij}$ and a determinant $h=\det h_{ij}$. In fact this is the main condition for possible solutions of the studied model:
\begin{proposition}
\emph{Integrability of the wave functional $\Psi[\xi]$ in the sense of functional integration in (\ref{norm}) determines the new dimension $\xi[h]$}.
\end{proposition}
\noindent The generalized dimension $\xi[h]$ can be established in the region of integrability $\Omega(h_I,h)$ as $\xi_{\Omega}[h_I,h]$ by using of the formula
\begin{equation}
  \xi_{\Omega}[h_I,h]=\int_{\Omega(h_I,h)}\delta \xi[h].
\end{equation}
In this paper we will study few consequences of the simplest choice (\ref{lg}). We will use standard argument which states that the normalization condition (\ref{norm}) establishes integrability constants of an arbitrary solution of the Schr\"odinger theory. The model in the invariant global dimension was discussed in detail in the Ref. \cite{glinka1}.

\section{Newton--Coulomb Potential\index{Newton--Coulomb potential}}\label{sec:2}

\subsection{Residual Approximation}
Let us consider the situation in which the series coefficients of the effective gravitational potential\index{gravitational potential!effective} are
\begin{equation}
  a_n=\left\{\begin{array}{cc}a_{-1}=const&\mathrm{for}~n=-1\\0&\mathrm{for}~n\neq-1\end{array}\right..
\end{equation}
We shall call such a case \emph{the residual approximation}\index{residual approximation}. In this approximation the effective gravitational potential\index{gravitational potential!effective} (\ref{eff}) takes the form
\begin{equation}
  V_{eff}=\dfrac{a_{-1}}{h-h_0},
\end{equation}
which is formally the Newton--Coulomb potential\index{Newton--Coulomb potential}, \emph{i.e.} has the behaviour like $1/h$ where $h$ is interpreted as a kind of radial quantity. The value of the coefficient $a_{-1}$ is unknown, but it is assumed that this coefficient exists. It can be verified by straightforward easy calculation that in such a situation the combined series coefficients $b_n$ are
\begin{equation}
b_n=\left\{\begin{array}{cc}
b_{-1}=h_0a_{-1}&\mathrm{for}~n=-1\\
b_{0}=a_{-1}&\mathrm{for}~n=0\\0&\mathrm{for}~n\neq-1,0\end{array}\right.,
\end{equation}
and, consequently, the Ricci scalar curvature\index{Ricci scalar curvature} of the three-dimensional space becomes
\begin{equation}\label{ric}
  {^{(3)}\!R}=2\Lambda+2\kappa \varrho-6\kappa^2a_{-1}\left(1+\dfrac{h_0}{h-h_0}\right),
\end{equation}
whereas the equation (\ref{kgf}), defining the global one-dimensional model of quantum gravity. takes the form
\begin{equation}\label{eq1a}
\left(\dfrac{\delta^2}{\delta h^2}+\dfrac{a_{-1}}{h-h_0}\right)\Psi=0.
\end{equation}
The Ricci scalar curvature\index{Ricci scalar curvature} (\ref{ric}) defines certain states of the geometry of the three-dimensional embedded space. However, even when one considers the case of vacuum, \emph{i.e.} when both the energy density\index{energy density} of Matter fields and cosmological constant\index{cosmological constant} are identically vanishing $\varrho=0$, $\Lambda=0$, it is rather difficult to establish a three dimensional metric tensor for which the Ricci scalar curvature\index{Ricci scalar curvature} behaves like
\begin{equation}
{^{(3)}\!R}\sim1+\dfrac{h_0}{h-h_0}.\label{rich}
\end{equation}

Interestingly, in the most general situation the residue of the three-dimensional Ricci scalar curvature\index{Ricci scalar curvature} calculated at a fixed point $h_0$ is
\begin{equation}
  \mathrm{Res}({^{(3)}}R,h_0)=2\kappa \mathrm{Res}(\varrho,h_0)-6\kappa^2a_{-1}h_0,
\end{equation}
\emph{i.e.} it can be taken \emph{ad hoc} identical to zero if and only if the residue of the energy density\index{energy density} of Matter fields\index{Matter fields} is
\begin{equation}
\mathrm{Res}(\varrho,h_0)=3\kappa a_{-1}h_0.
\end{equation}
If one takes \emph{ad hoc} the following relation
\begin{equation}
a_{-1}=\dfrac{\Lambda}{3\kappa^2},
\end{equation}
then the Ricci scalar curvature\index{Ricci scalar curvature} of an induced three-dimensional geometry of  embedded space takes the form
\begin{equation}\label{ric}
  {^{(3)}\!R}=2\kappa \varrho-\dfrac{2\Lambda{h_0}}{h-h_0},
\end{equation}
and its residue at a fixed point $h_0$
\begin{equation}
  \mathrm{Res}({^{(3)}}R,h_0)=2\kappa \mathrm{Res}(\varrho,h_0)-2\Lambda{h_0},
\end{equation}
identically vanishes if and only if the residue of the energy density\index{energy density} of Matter fields at a fixed point $h_0$ has the value
\begin{equation}
\mathrm{Res}(\varrho,h_0)=\dfrac{\Lambda}{\kappa }h_0.
\end{equation}
In such a situation, also the geometry of an embedded three-dimensional space manifold is automatically Ricci-flat\index{Ricci-flat manifold} if and only if the energy density\index{energy density} of Matter fields\index{Matter fields} takes the following form
\begin{equation}
  \varrho=\dfrac{\Lambda}{\kappa }\dfrac{{h_0}}{h-h_0}.
\end{equation}
It is easy to see that another possible Ricci-flat three-dimensional manifold\index{Ricci-flat manifold} is obtained for the identically vanishing cosmological constant\index{cosmological constant} $\Lambda=0$ and the following value of the energy density\index{energy density} of Matter fields
\begin{eqnarray}
\varrho=3\kappa a_{-1}\left(1+\dfrac{h_0}{h-h_0}\right).
\end{eqnarray}

\subsection{Newton--Coulomb Quantum Gravity}
In the most general situation three-dimensional spaces having induced metric\index{induced metric}s characterized by the Ricci scalar curvature\index{Ricci scalar curvature} of the form (\ref{rich}) are not yet known explicitly in literature. However, it is evident that in the particular situation $h_0=0$, which is fully consistent with the general condition (\ref{h0cond}), the state of affairs is determined by much more simplified equations
\begin{eqnarray}
&&{^{(3)}\!R}=2\Lambda+2\kappa \varrho-6\kappa^2a_{-1},\label{ncqg1}\\
&&\left(\dfrac{\delta^2}{\delta h^2}+\dfrac{a_{-1}}{h}\right)\Psi=0.\label{ncqg2}
\end{eqnarray}
Let us consider this particular case as the basic situation. We shall call the global one-dimensional quantum gravity\index{global one-dimensional quantum gravity} described by the system of equations (\ref{ncqg1})-(\ref{ncqg2}) \emph{the Newton--Coulomb quantum gravity}.\index{Newton--Coulomb quantum gravity}

As an example, we shall consider first the case of constant Ricci curvature. It is not difficult to see that in the most general view such a situation corresponds to the identically vanishing energy density\index{energy density} of Matter fields
\begin{equation}\label{kgf3}
\varrho\equiv 0.
\end{equation}
We shall call such a case \emph{the Newton--Coulomb stationary quantum gravity}\index{Newton--Coulomb quantum gravity!stationary}. In such particular situation, the Ricci scalar curvature\index{Ricci scalar curvature} of three-dimensional embedded space becomes
\begin{equation}
  {^{(3)}}R=2\Lambda-6\kappa^2a_{-1}=const,
\end{equation}
and, consequently, the Ricci curvature tensor\index{Ricci curvature tensor}, which characterizes the intrinsic geometry of the manifolds, describes the three-dimensional Einstein manifolds\index{Einstein manifolds} \cite{besse}
\begin{equation}\label{eins}
  R_{ij}=\lambda h_{ij},
\end{equation}
where the sign $\lambda$ of the Einstein manifolds\index{Einstein manifolds}, in these specific conditions, is completely defined by parameters of the Newton--Coulomb stationary quantum gravity\index{Newton--Coulomb quantum gravity!stationary} -- the cosmological constant\index{cosmological constant} and the residue of the effective gravitational potential\index{gravitational potential!effective} -- as follows
\begin{equation}\label{sign}
  \dfrac{2}{3}\Lambda-2\kappa^2a_{-1}=\lambda.
\end{equation}
Interestingly, the crucial consequence of identical vanishing of the energy density\index{energy density} of Matter fields is the property of \emph{maximal symmetry} of the Einstein manifolds\index{Einstein manifolds} described by the sign (\ref{sign}). For this reason, the Newton--Coulomb stationary quantum gravity\index{Newton--Coulomb quantum gravity!stationary} possesses highly non-trivial geometrical interpretation: such a situation describes embedded three-dimensional manifolds which are \emph{maximally symmetric Einstein manifolds}.\index{Einstein manifolds}

Consequently, one can deduce straightforwardly the classification of the three-dimensional embedded spaces, which are maximally symmetric three-dimensional Einstein manifolds\index{Einstein manifolds} (\ref{eins}), with respect to the value of the sign $\lambda$ (\ref{sign}) of a manifold in dependence on the cosmological constant\index{cosmological constant} and the residue $a_{-1}$:

\begin{conclusion}[Classification of maximally symmetric three-dimensional Einstein manifolds]
The Newton--Coulomb stationary quantum gravity\index{Newton--Coulomb quantum gravity!stationary}, defined by the effective gravitational potential\index{gravitational potential!effective} $V_{eff}[h]=\dfrac{a_{-1}}{h}$, determines the three-dimensional embedded spaces which are the maximally symmetric three-dimensional Einstein manifolds\index{Einstein manifolds}, characterized by the sign of the form (\ref{sign}). There is the classification of such manifolds with respect to the cosmological constant\index{cosmological constant} $\Lambda$ and the value $a_{-1}$ of the residue of the effective gravitational potential\index{gravitational potential!effective}
\begin{enumerate}
\item If the sign of manifold is non-zero $\lambda\neq0$ and the residue of the effective gravitational potential\index{gravitational potential!effective} is a negative real $a_{-1}=-|\alpha|$, then the effective gravitational potential\index{gravitational potential!effective} $V_{eff}[h]$ corresponds to the Newtonian attractive potential energy
    \begin{equation}
    V_{eff}=-\dfrac{|\alpha|}{h}\sim-\dfrac{m_1m_2}{h}.\label{i1}
    \end{equation}
    \begin{enumerate}
      \item If the cosmological constant\index{cosmological constant} is a positive real $\Lambda=+|\Lambda|$ then the maximally symmetric Einstein three-manifolds are characterized by the positive Ricci scalar curvature\index{Ricci scalar curvature}
    \begin{equation}
    {^{(3)}}R=\dfrac{2}{3}|\Lambda|+2\kappa^2|\alpha|.
    \end{equation}
    \item If the cosmological constant\index{cosmological constant} is a negative real $\Lambda=-|\Lambda|$ then the maximally symmetric Einstein three-manifolds are characterized by the Ricci scalar curvature\index{Ricci scalar curvature}
    \begin{equation}
    {^{(3)}}R=-\dfrac{2}{3}|\Lambda|+2\kappa^2|\alpha|,
    \end{equation}
    which is
    \begin{equation}
      {^{(3)}}R\left\{\begin{array}{ccc}
        <0,&\textrm{iff}&|\Lambda|>3\kappa^2|\alpha|\\
        =0,&\textrm{iff}&|\Lambda|=\kappa^2|\alpha|\\
        >0,&\textrm{iff}&|\Lambda|<3\kappa^2|\alpha|
      \end{array}\right.
    \end{equation}
    \end{enumerate}
\item If the sign of manifold is non-zero $\lambda\neq0$ and the residue of the effective gravitational potential\index{gravitational potential!effective} is a positive real $a_{-1}=+|\alpha|$, then the effective potential $V_{eff}[h]$ becomes the Coulomb repulsive potential energy
    \begin{equation}
    V_{eff}=\dfrac{|\alpha|}{h}\sim\dfrac{q_1q_2}{h}.\label{i2}
    \end{equation}
    \begin{enumerate}
      \item If the cosmological constant\index{cosmological constant} is a negative real $\Lambda=-|\Lambda|$ then the maximally symmetric Einstein three-manifolds are characterized by negative Ricci scalar curvature\index{Ricci scalar curvature}
    \begin{equation}
      {^{(3)}}R=-\dfrac{2}{3}|\Lambda|-2\kappa^2|\alpha|.
    \end{equation}
    \item If the cosmological constant\index{cosmological constant} is a positive real $\Lambda=+|\Lambda|$ then the maximally symmetric Einstein three-manifolds are characterized by the Ricci scalar curvature\index{Ricci scalar curvature}
    \begin{equation}
      {^{(3)}}R=\dfrac{2}{3}|\Lambda|-2\kappa^2|\alpha|,
    \end{equation}
    which is
    \begin{equation}
      {^{(3)}}R\left\{\begin{array}{ccc}
        <0,&\textrm{iff}&|\Lambda|<3\kappa^2|\alpha|\\
        =0,&\textrm{iff}&|\Lambda|=\kappa^2|\alpha|\\
        >0,&\textrm{iff}&|\Lambda|>3\kappa^2|\alpha|
      \end{array}\right.
    \end{equation}
    \end{enumerate}
\item If the sign of manifold is identically vanishing $\lambda=0$, \emph{i.e.} the maximally symmetric Einstein three-manifolds\index{Einstein manifolds} are Ricci-flat\index{Ricci-flat manifold} manifolds, then one can determine uniquely the value of the residue of the effective gravitational potential\index{gravitational potential!effective} as follows
    \begin{equation}
    a_{-1}=\pm\dfrac{|\Lambda|}{3\kappa^2}.\label{i3}
    \end{equation}
    In such a situation, one obtains the values of the cosmological constant\index{cosmological constant}
\begin{equation}\label{lambda}
  |\Lambda|\sim\left\{\begin{array}{rl}r_g(m_1)r_g(m_2)&\mathrm{for~the~Newton~law}\vspace*{10pt}\\
  r_e(q_1)r_e(q_2)&\mathrm{for~the~Coulomb~law}\end{array}\right.
\end{equation}
where $m$ is the mass of a body generating Newtonian gravitational field in vacuum and $r_g(m)=\dfrac{2Gm}{c^2}$ is its gravitational radius, $q$ is the charge generating Coulombic electrical field in vacuum and $r_e(q)=q\sqrt{\dfrac{G}{4\pi \epsilon_0c^4}}$ is its electrical radius.
\end{enumerate}
\end{conclusion}

Note that, in fact, by taking into account \emph{ad hoc} the relation for the series coefficients (\ref{calka}), the residue $a_{-1}$ of the effective gravitational potential\index{gravitational potential!effective} is the Cauchy integral of $V_{eff}$ at a fixed point $h_0=0$
\begin{equation}
  a_{-1}=\textrm{Res}\left[\dfrac{1}{6\kappa^2}\dfrac{1}{h}\left(-{^{(3)}\!R}+2\Lambda+2\kappa \varrho\right),h=0\right],
\end{equation}
and its value can be straightforwardly established as
\begin{eqnarray}
  a_{-1}&=&\dfrac{1}{6\kappa^2}\left.\left(-{^{(3)}\!R}+2\Lambda+2\kappa \varrho\right)\right|_{h=0}\nonumber\\
  &=&-\dfrac{{^{(3)}}R_0}{6\kappa^2}+\dfrac{\Lambda}{3\kappa^2}+\dfrac{\kappa \varrho_0}{3\kappa^2},\label{resi}
\end{eqnarray}
where subscript “$0$” means the value of a quantity calculated in $h=0$.

Let us note that, when one shall to associate the residual ef\/fective potential $V_{eff}[h]=\dfrac{a_{-1}}{h}$ with any realistic quantized Kepler problem, i.e. with employing the Newtonian or the Coulombic potentials, one should to identify the global dimension with a spatial distance $r=\sqrt{x^2+y^2+z^2}$
\begin{equation}\label{detr}
  h\equiv r\quad,
\end{equation}
In this case, with the formal identif\/ication of the functional derivative of $h$ and the classical derivative of $r$, \emph{i.e.} in fact the equality between the functional and classical integral measures $\delta h = dr$ well--known in classical mechanics \cite{gold}, The wave functional $\Psi[h]$ becomes the radial wave function $\Psi(r)$, and the evolution (\ref{eq1a}) becomes familiar radial type Schr\"odinger equation
\begin{equation}\label{eq1ab}
  \left(\dfrac{d^2}{d{r^2}}+\dfrac{\mp|\alpha|}{r}\right)\Psi(r)=0\quad,
\end{equation}
where the number $|\alpha|$ can be taken straightforwardly from the Newton law of gravitation or from the Coulomb law of electricity. The received wave equation (\ref{eq1ab}) possibly describes an atomic system.

Note that there is many possible choices of the metrics $h_{ij}$ with the same value of the determinant $h=r$. For instance one can take the simple variant
\begin{eqnarray}
  h_{ij}=r^{1/3}\delta_{ij}.
\end{eqnarray}
However, more generally, one can parameterize the relation (\ref{detr}) by $SO(3)$ group rotation matrix $r_{ij}$: $h_{ij}=r^{1/3}r_{ij}$, which allows use the Eulerian angles $(\theta,\varphi,\phi)$ as follows
\begin{equation}
  r_{ij}(\theta,\varphi,\phi)\equiv r_{il}^{(3)}(\theta)r_{lk}^{(2)}(\varphi)r_{kj}^{(3)}(\phi)\quad,
\end{equation}
where matrices $r_{ij}^{(p)}(\vartheta)$ are rotation matrices around the selected $p$-axis
\begin{eqnarray}
r_{ij}^{(3)}(\vartheta)=\left[\begin{array}{ccc}\cos\vartheta&-\sin\vartheta&0\\
\sin\vartheta&\cos\vartheta&0\\
0&0&1\end{array}\right],\qquad r_{ij}^{(2)}(\vartheta)=\left[\begin{array}{ccc}\cos\vartheta&0&\sin\vartheta\\
0&1&0\\
-\sin\vartheta&0&\cos\vartheta\end{array}\right].
\end{eqnarray}
This point of view was discussed in much advanced detail in the Ref. \cite{glinka1}.

\section{Geometric wave functionals}\label{sec:3}
In this section we shall consider certain solutions of the global one-dimensional model of quantum gravity (\ref{kgf}) for the case of the residual approximation of the effective gravitational potential $V_{eff}$ implemented in the previous section. In the most general situation, the considered quantum mechanical evolution
\begin{equation}\label{kgf5}
\left(\dfrac{\delta^2}{\delta{h^2}}\mp\dfrac{|\alpha|}{h}\right)\Psi^{\mp}[h]=0,
\end{equation}
is solved by two types of wave functions $\Psi^{\mp}$ where the attractive wave functions $\Psi_G^{-}[h]$ are associated with the Newton-like effective gravitational potential, and the repulsive ones $\Psi^{+}[h]$ are associated with the Coulomb-like effective gravitational potential. Because of the manifest one-dimensionality of the functional evolutionary equation (\ref{kgf5}), one can solve this equation in the framework of the theory of ordinary differential equations by interpretation of the functional derivative as the ordinary one, \emph{i.e.} $\dfrac{\delta}{\delta h} = \dfrac{d}{dh}$, and the wave functional as a wave function $\Psi[h]=\Psi(h)$ with no loss of generality.

In this manner, the problem to solve is given by the second order ordinary differential equation
\begin{equation}\label{kgf6}
\left(\dfrac{d^2}{dh^2}\mp\dfrac{|\alpha|}{h}\right)\Psi^{\mp}(h)=0,
\end{equation}
which is well-known in the mathematical physics literature. The general solution of such a differential equation can be constructed straightforwardly by making use of the Bessel functions $J_n$ and $Y_n$ for the case of the attractive potential
\begin{equation}
  \Psi_G^{-}[h]=\sqrt{|\alpha| h}\left[C_1^-J_1\left(2\sqrt{|\alpha| h}\right) + 2iC_2^-Y_1\left(2\sqrt{|\alpha| h}\right)\right]\quad,\label{solp}
\end{equation}
and in terms of the modif\/ied Bessel functions $I_n$ and $K_n$ for the case of repulsive potential
\begin{equation}
\Psi_G^{+}[h]=-\sqrt{|\alpha| h}\left[C_1^+I_1\left(2\sqrt{|\alpha| h}\right) + 2C_2^+K_1\left(2\sqrt{|\alpha| h}\right)\right]\quad,\label{solm}
\end{equation}
where $C_1^{\pm}$ and $C_2^{\pm}$ are constants of integration, one takes standard def\/initions \cite{spec} of the Bessel functions of f\/irst and second kind, $J_\alpha(x)$ and $Y_\alpha(x)$,
\begin{eqnarray}
   J_\alpha(x)&=&\dfrac{1}{\pi}\int_0^\pi dt\cos\left(x\cos t-\alpha t\right)\quad,\\
  Y_\alpha(x)&=&\dfrac{J_\alpha(x)\cos\left(\alpha\pi\right)-J_{-\alpha}(x)}{\sin\left(\alpha\pi\right)}\quad,
\end{eqnarray}
and the modif\/ied Bessel functions of f\/irst and second kind, $I_\alpha(x)$ and $K_\alpha(x)$,
\begin{eqnarray}
   I_\alpha(x)&=&\dfrac{1}{\pi}\int_0^\pi dt\exp\left(x\cos t\right)\cos\left(\alpha t\right)\quad,\\
  K_\alpha(x)&=&\dfrac{\pi}{2}\dfrac{I_{-\alpha}(x)-I_\alpha(x)}{\sin\left(\alpha \pi\right)}.
\end{eqnarray}
Standardly, values of the second kind Bessel functions and modif\/ied ones for any integers $n$ can be received by application of the limiting procedure $Y_n(x)=\lim_{\alpha \rightarrow n}Y_\alpha(x)$, $K_n(x)=\lim_{\alpha \rightarrow n}K_\alpha(x)$ .

In further parts of this section we shall to present solutions of the quantum mechanics (\ref{kgf5}) with respecting of few selected boundary conditions for the general solutions (\ref{solp}) and (\ref{solm}).

\subsection{Boundary conditions I}\label{sub:4}
Let us consider the global one-dimensional quantum mechanics (\ref{kgf}) with the boundary conditions for some selected initial value of the dimension $h=h_I$:
\begin{equation}\label{bound1}
  \Psi[h_I]=\Psi_I\quad,\quad\dfrac{\delta\Psi}{\delta h}[h_I]=\Psi'_I.
\end{equation}
With using of the regularized hypergeometric functions ${_p}\tilde{F}_q$
\begin{eqnarray}
{_p}\tilde{F}_q\left(\begin{array}{c} a_1,\ldots,a_p\\b_1,\ldots,b_q \end{array};x\right)&=&\dfrac{{_p}F_q\left(\begin{array}{c} a_1,\ldots,a_p\\b_1,\ldots,b_q \end{array};x\right)}{\Gamma(b_1)\ldots\Gamma(b_q)},\\
{_p}F_q\left(\begin{array}{c} a_1,\ldots,a_p\\b_1,\ldots,b_q \end{array};x\right)&=&\sum_{r=0}^{\infty}\dfrac{(a_1)_r\ldots(a_p)_r}{(b_1)_r\ldots(b_q)_r}\dfrac{x^r}{r!},\\
(a)_r&\equiv&\dfrac{\Gamma(a+r)}{\Gamma(a)},
\end{eqnarray}
one can write out the general solutions (\ref{solp}) and (\ref{solm}) with respect to the boundary conditions (\ref{bound1})
\begin{eqnarray}\label{sol1}
\Psi_G^{-}=C^-_1\left(2\sqrt{\strut{|\alpha| h}}\right)K_1\left(2\sqrt{\strut{|\alpha| h}}\right)+C^-_2\left(2\sqrt{\strut{|\alpha| h}}\right)^2{_0}\tilde{F}_1\left(\begin{array}{c} -\\2\end{array};|\alpha| h\right),
\end{eqnarray}
where the sign $-$ in the hypergeometric function notation means that all $(a)_r=1$, and the constans
\begin{eqnarray}
  C^-_1&=&\Psi_I\,{_0}\tilde{F}_1\left(\begin{array}{c} -\\1\end{array};|\alpha| h_I\right)-\Psi'_Ih_I\,{_0}\tilde{F}_1\left(\begin{array}{c} -\\2\end{array};|\alpha| h_I\right),\\
  C^-_2&=&\dfrac{1}{2}\left(\Psi_IK_0\left(2\sqrt{\strut{|\alpha| h_I}}\right)+\Psi'_I\sqrt{\strut{\dfrac{h_I}{|\alpha|}}}K_1\left(2\sqrt{\strut{|\alpha| h_I}}\right)\right),
\end{eqnarray}
for Newtonian case, and
\begin{eqnarray}\label{sol2}
\Psi_G^{+}=C^+_1\left(2\sqrt{\strut{|\alpha| h}}\right)Y_1\left(2\sqrt{\strut{|\alpha| h}}\right)+C^+_2\left(2\sqrt{\strut{|\alpha| h}}\right)^2{_0}\tilde{F}_1\left(\begin{array}{c} -\\2\end{array};-|\alpha| h\right),
\end{eqnarray}
with constans
\begin{eqnarray}
  C^+_1&=&\dfrac{\pi}{2}\left(\Psi'_Ih_I\,{_0}\tilde{F}_1\left(\begin{array}{c} -\\2\end{array};-|\alpha| h_I\right)-\Psi_I\,{_0}\tilde{F}_1\left(\begin{array}{c} -\\1\end{array};-|\alpha| h_I\right)\right),\\
  C^+_2&=&\dfrac{\pi}{2}\left(\Psi_IY_0\left(2\sqrt{\strut{|\alpha| h_I}}\right)-\Psi'_I\sqrt{\strut{\dfrac{h_I}{|\alpha|}}}Y_1\left(2\sqrt{\strut{|\alpha| h_I}}\right)\right),
\end{eqnarray}
for Coulombic case.

\subsection{Boundary conditions II}\label{sub:5}
The second case which we want to present in this paper, are the boundary conditions for 1st and 2nd functional derivatives
\begin{equation}\label{bound2}
  \dfrac{\delta\Psi}{\delta h}[h_I]=\Psi'_I\quad,\quad\dfrac{\delta^2\Psi}{\delta h^2}[h_I]=\Psi''_I.
\end{equation}
By using of the hypergeometric functions, one can express the solution for attractive case as follows
\begin{eqnarray}\label{sol1a}
\Psi_G^{-}=C^-_1\left(2\sqrt{\strut{|\alpha| h}}\right)K_1\left(2\sqrt{\strut{|\alpha| h}}\right)+C^-_2\left(2\sqrt{\strut{|\alpha| h}}\right)^2{_0}\tilde{F}_1\left(\begin{array}{c} -\\2\end{array};|\alpha| h\right),
\end{eqnarray}
where $C^-_1$ and $C^-_2$ are constants def\/ined as
\begin{eqnarray}
  C^-_1&=&-h_I\left(\Psi'_I\,{_0}\tilde{F}_1\left(\begin{array}{c} -\\2\end{array};|\alpha| h_I\right)-\dfrac{\Psi''_I}{|\alpha|}\,{_0}\tilde{F}_1\left(\begin{array}{c} -\\1\end{array};|\alpha| h_I\right)\right),\\
  C^-_2&=&\dfrac{1}{2}\sqrt{\dfrac{h_I}{|\alpha|}}\left(\Psi''_I\sqrt{\dfrac{h_I}{|\alpha|}}K_0\left(2\sqrt{\strut{|\alpha| h_I}}\right)+\Psi'_IK_1\left(2\sqrt{\strut{|\alpha| h_I}}\right)\right).
\end{eqnarray}
Similarly for the repulsive case one obtains easily
\begin{eqnarray}\label{sol2a}
\Psi_G^{+}=C^+_1\left(2\sqrt{\strut{|\alpha| h}}\right)Y_1\left(2\sqrt{\strut{|\alpha| h}}\right)+C^+_2\left(2\sqrt{\strut{|\alpha| h}}\right)^2{_0}\tilde{F}_1\left(\begin{array}{c} -\\2\end{array};-|\alpha| h\right),
\end{eqnarray}
with constans
\begin{eqnarray}
  C^+_1&=&\dfrac{\pi h_I}{2}\left(\Psi'_I\,{_0}\tilde{F}_1\left(\begin{array}{c} -\\2\end{array};-|\alpha| h_I\right)+\dfrac{\Psi''_I}{|\alpha|}\,{_0}\tilde{F}_1\left(\begin{array}{c} -\\1\end{array};-|\alpha| h_I\right)\right),\\
  C^+_2&=&\dfrac{\pi}{4}\sqrt{\dfrac{h_I}{|\alpha|}}\left(\Psi''_I\sqrt{\dfrac{h_I}{|\alpha|}}Y_0\left(2\sqrt{\strut{|\alpha| h_I}}\right)+\Psi'_IY_1\left(2\sqrt{\strut{|\alpha| h_I}}\right)\right).
\end{eqnarray}

\subsection{Boundary conditions III}\label{sub:6}
The last possible case of boundary conditions for the considered problem is
\begin{equation}\label{bound3}
  \Psi[h_I]=\Psi_I\quad,\quad\dfrac{\delta^2\Psi}{\delta h^2}[h_I]=\Psi''_I.
\end{equation}
These conditions are formally improper for the problem, because of they lead to singular solutions. In this case, however, one can present the solutions in the form with formally singular constans. For the attractive potential one has
\begin{equation}\label{sol1c}
  \Psi_G^-=C^-_1\left(2\sqrt{|\alpha|h}\right)K_1\left(2 \sqrt{\left|\alpha\right|h}\right)+C^-_2\left(2\sqrt{|\alpha|h}\right)^2{_0}\tilde{F}_1\left(\begin{array}{c} -\\2\end{array};|\alpha| h\right),
\end{equation}
with constans ($\epsilon\rightarrow0$)
\begin{eqnarray}
  C^-_1&=&\dfrac{2}{\epsilon}\sqrt{|\alpha|h_I}\left(\Psi_I-\dfrac{h_I}{\left|\alpha\right| }\Psi''_I \right){_0}\tilde{F}_1\left(\begin{array}{c} -\\2\end{array};|\alpha| h_I\right),\\
  C^-_2&=&\dfrac{1}{\epsilon}\left(\Psi_I-\dfrac{h_I}{|\alpha|}\Psi''_I\right)K_1\left(2\sqrt{|\alpha|h_I}\right),
\end{eqnarray}
and similarly for the repulsive potential one obtains
\begin{equation}\label{sol2c}
  \Psi_G^+=C^+_1\left(2\sqrt{|\alpha|h}\right)Y_1\left(2 \sqrt{\left|\alpha\right|h}\right)+C^+_2\left(2\sqrt{|\alpha|h}\right)^2{_0}\tilde{F}_1\left(\begin{array}{c} -\\2\end{array};-|\alpha| h\right),
\end{equation}
with constants ($\epsilon\rightarrow0$)
\begin{eqnarray}
  C^+_1&=&\dfrac{2}{\epsilon}\sqrt{|\alpha|h_I}\left(\Psi_I+\dfrac{h_I}{\left|\alpha\right| }\Psi''_I \right){_0}\tilde{F}_1\left(\begin{array}{c} -\\2\end{array};-|\alpha| h_I\right),\\
  C^+_2&=&\dfrac{1}{\epsilon}\left(\Psi_I+\dfrac{h_I}{|\alpha|}\Psi''_I\right)Y_1\left(2\sqrt{|\alpha|h_I}\right).
\end{eqnarray}
However, when the following relation for the boundary conditions holds
\begin{equation}\label{cond}
  \pm\dfrac{h_I}{\left|\alpha\right| }{\Psi^{\pm}_I}''+\Psi^{\pm}_I\equiv\epsilon f_{\pm}[h_I,|\alpha|],
\end{equation}
where $f_{\pm}[h_I,|\alpha|]\neq0$ are some (now unknown and arbitrary) nonsingular functionals of $h_I$ and $|\alpha|$, the sign $+$ is related to the Newtonian case, and the sign $-$ to the Coulombic one, then the singularity of the solutions (\ref{sol1c}) and (\ref{sol2c}) is canceling. In this case the initial value $\Psi_I$ for the attractive case is
\begin{eqnarray}
\!\!\!\!\!\!\!\!\!\!&&\Psi^-_I=-|\alpha|h_I\,{_0}{F}_1\left(\begin{array}{c} -\\2\end{array};|\alpha|h_I\right)\left[c^-_1+2\epsilon\sqrt{|\alpha|}\int_1^{h_I}\dfrac{dt}{\sqrt{t}}f_{-}[t,|\alpha|]K_1\left(2\sqrt{|\alpha|t}\right)\right]+\nonumber\\
\!\!\!\!\!\!\!\!\!\!&&+\,2\sqrt{|\alpha|h_I}K_1\left(2\sqrt{|\alpha|h_I}\right)\left[c^-_2+\epsilon|\alpha|\int_1^{h_I}dtf_{-}[t,|\alpha|]\,{_0}{F}_1\left(\begin{array}{c} -\\2\end{array};|\alpha|t\right)\right],\label{eq1}
\end{eqnarray}
and similarly for the repulsive one
\begin{eqnarray}
\!\!\!\!\!\!\!\!\!\!&&\Psi^+_I=|\alpha|h_I\,{_0}{F}_1\left(\begin{array}{c} -\\2\end{array};-|\alpha|h_I\right)\left[c^+_1-\epsilon\pi\sqrt{|\alpha|}\int_1^{h_I}\dfrac{dt}{\sqrt{t}}f_{+}[t,|\alpha|]Y_1\left(2\sqrt{|\alpha|t}\right)\right]+\nonumber\\
\!\!\!\!\!\!\!\!\!\!&&+\,2i\sqrt{|\alpha|h_I}Y_1\left(2\sqrt{|\alpha|h_I}\right)\left[c^+_2-\epsilon\dfrac{i\pi}{2}|\alpha|\int_1^{h_I}dtf_{+}[t,|\alpha|]\,{_0}{F}_1\left(\begin{array}{c} -\\2\end{array};-|\alpha|t\right)\right],\label{eq2}
\end{eqnarray}
where $c^{\pm}_{1,2}$ are nonsingular constants of integration. The functionals $f_{\pm}[h_I,|\alpha|]\neq0$ can be established by application of the condition (\ref{cond}) within the general solutions (\ref{sol1c}) and (\ref{sol2c}). It yields the results
\begin{eqnarray}
  \Psi_I^-&=&8|\alpha|h_IK_1\left(2\sqrt{|\alpha|h_I}\right)\,{_0}\tilde{F}_1\left(\begin{array}{c} -\\2\end{array};|\alpha|h_I\right)f_{-}[h_I,|\alpha|],\\
  \Psi_I^+&=&8|\alpha|h_IY_1\left(2\sqrt{|\alpha|h_I}\right)\,{_0}\tilde{F}_1\left(\begin{array}{c} -\\2\end{array};-|\alpha|h_I\right)f_{+}[h_I,|\alpha|].
\end{eqnarray}
Employing straightforwardly these results into the equalities (\ref{eq1}) and (\ref{eq2}) one obtains the integral equations for the functionals $f_{\pm}$. For the Coulombic situation one receives the following equation
\begin{eqnarray}
&&-\dfrac{c^-_1}{4}\left(2\sqrt{|\alpha|h_I}\right)\,{_0}{F}_1\left(\begin{array}{c} -\\2\end{array};|\alpha|h_I\right)+c^-_2K_1\left(2\sqrt{|\alpha|h_I}\right)+\nonumber\\
&+&\epsilon|\alpha|\Bigg\{K_1\left(2\sqrt{|\alpha|h_I}\right)\int_1^{h_I}dt\,{_0}{F}_1\left(\begin{array}{c} -\\2\end{array};|\alpha|t\right)-\nonumber\\
&-&\sqrt{h_I}\,{_0}{F}_1\left(\begin{array}{c} -\\2\end{array};|\alpha|h_I\right)\int_1^{h_I}\dfrac{dt}{\sqrt{t}}K_1\left(2\sqrt{|\alpha|t}\right)\Bigg\}f_{-}[t,|\alpha|]=\nonumber\\
&=&2\left(2\sqrt{|\alpha|h_I}\right)K_1\left(2\sqrt{|\alpha|h_I}\right)\,{_0}\tilde{F}_1\left(\begin{array}{c} -\\2\end{array};|\alpha|h_I\right)f_{-}[h_I,|\alpha|],\label{93}
\end{eqnarray}
and similarly for the Newtonian case one derives the relation
\begin{eqnarray}
&&\dfrac{c^+_1}{4}\left(2\sqrt{|\alpha|h_I}\right)\,{_0}{F}_1\left(\begin{array}{c} -\\2\end{array};-|\alpha|h_I\right)+ic^+_2Y_1\left(2\sqrt{|\alpha|h_I}\right)-\nonumber\\
&-&\epsilon\dfrac{\pi}{2}|\alpha|\Bigg\{\sqrt{h_I}\,{_0}{F}_1\left(\begin{array}{c}-\\2\end{array};-|\alpha|h_I\right)\int_1^{h_I}\dfrac{dt}{\sqrt{t}}Y_1\left(2\sqrt{|\alpha|t}\right)+\nonumber\\
&+&\,iY_1\left(2\sqrt{|\alpha|h_I}\right)\int_1^{h_I}dt\,{_0}{F}_1\left(\begin{array}{c} -\\2\end{array};-|\alpha|t\right)\Bigg\}f_{+}[t,|\alpha|]=\nonumber\\
&=&2\left(2\sqrt{|\alpha|h_I}\right)Y_1\left(2\sqrt{|\alpha|h_I}\right)\,{_0}\tilde{F}_1\left(\begin{array}{c} -\\2\end{array};-|\alpha|h_I\right)f_{+}[h_I,|\alpha|].\label{94}
\end{eqnarray}
In both the cases the integral operators acting on the functionals $f_{\pm}$ are nonsingular. By this reason one can put straightforwardly the formal limit $\epsilon\rightarrow0$ in the integral equations (\ref{93}) and (\ref{94}), and by doing few elementary algebraic manipulations one can extract the searched functionals. The f\/inal results are as follows
\begin{eqnarray}
f_{-}[h_I,|\alpha|]&=&\dfrac{-c^-_1/8}{K_1\left(2\sqrt{|\alpha|h_I}\right)}+\dfrac{c^-_2/4}{I_1\left(2\sqrt{|\alpha|h_I}\right)},\\
f_{+}[h_I,|\alpha|]&=&\dfrac{c^+_1/8}{Y_1\left(2\sqrt{|\alpha|h_I}\right)}+\dfrac{ic^+_2/4}{J_1\left(2\sqrt{|\alpha|h_I}\right)}.
\end{eqnarray}
In this manner the initial data conditions for the considered boundary conditions (\ref{bound3}) can not be chosen arbitrary, but according to the rules
\begin{eqnarray}
  \Psi_I^-&=&\sqrt{|\alpha|h_I}\left[-c^-_1I_1\left(2\sqrt{|\alpha|h_I}\right)+2c^-_2K_1\left(2\sqrt{|\alpha|h_I}\right)\right],\\
  \Psi_I^+&=&\sqrt{|\alpha|h_I}\left[c^+_1J_1\left(2\sqrt{|\alpha|h_I}\right)+2ic^+_2Y_1\left(2\sqrt{|\alpha|h_I}\right)\right].
\end{eqnarray}

The supposed equation for boundary values (\ref{cond}) is not unique, and can be replaced by other conditions. The discussed case, however, reflects the typical questions arising within the problem.

\section{Discussion}\label{sec:5}

In this paper we have discussed the quantum mechanical model of quantum gravity arising from the global one--dimensional conjecture within quantum general relativity considered recently by the author \cite{glinka1}. This model straightforwardly bases on the ef\/fective potential (\ref{eff}) being a simple algebraic sum of three fundamental energetic constituents - ''geometric'', ''cosmological'', and ''material'', with nontrivial change in potential behavior with respect to the initial model that was the Wheeler--DeWitt quantum geometrodynamics (\ref{wdw}). The relation between the models is established by change of both the dif\/ferential operator and the potential.

We have considered the analytical form of the ef\/fective potential, and concentrated an especial attention on the physical conclusions following from the \emph{residual ef\/fective potential}, which on some well-established conventional level is directly identif\/ied with the attractive Newton's gravitation or the repulsive Coulomb's electrostatics. Studying of this special case allowed to conclude that in the case of Matter f\/ields's energy absence, in the global one--dimensional model of quantum gravity, the maximally symmetric three-dimensional Einstein manifolds are the characteristic embeddings for the residual ef\/fective potential. F\/inally, we have found some solutions of the model of quantum gravity in the residual approximation.

We hope to discuss further conclusions of the global-one dimensional model of quantum gravity in the next contributions.

\end{document}